\documentclass[aps,prb,twocolumn,superscriptaddress]{revtex4-1}

\renewcommand{\thefootnote}{\fnsymbol{footnote}}
\renewcommand{\theequation}{\arabic{section}.\arabic{equation}}

\usepackage{amssymb}
\usepackage{amsfonts}
\usepackage{amsmath}
\usepackage{bm}
\usepackage{textcomp}
\usepackage{color}
\usepackage{longtable}
\usepackage{graphicx}
\usepackage{dcolumn}

\usepackage{epsfig}

\begin{document}
\renewcommand{\thefootnote}{\fnsymbol{footnote}}
\renewcommand{\theequation}{\arabic{section}.\arabic{equation}}

\title{Phase diagram and critical behavior of the random ferromagnet Ga$_{1-x}$Mn$_x$N}

\author{S. Stefanowicz} \email{sstefanowicz@ifpan.edu.pl}
\affiliation{Institute of Physics, Polish Academy of Sciences, PL-02 668 Warszawa, Poland}

\author{G. Kunert}
\affiliation{Institute of Solid State Physics, University of Bremen, D-28359 Bremen, Germany}

\author{C. Simserides}
\affiliation{Physics Department, University of Athens, GR-15784 Athens, Greece}

\author{J. A. Majewski}
\affiliation{Institute of Theoretical Physics, Faculty of Physics, University of Warsaw, PL-00 681 Warszawa, Poland}

\author{W. Stefanowicz}
\affiliation{Institute of Physics, Polish Academy of Sciences, PL-02 668 Warszawa, Poland}

\author{C. Kruse}
\affiliation{Institute of Solid State Physics, University of Bremen, D-28359 Bremen, Germany}

\author{S. Figge}
\affiliation{Institute of Solid State Physics, University of Bremen, D-28359 Bremen, Germany}

\author{Tian Li}
\affiliation{Institut f\"ur Halbleiter- und Festk\"orperphysik, Johannes Kepler University,  A-4040 Linz, Austria}

\author{R. Jakie{\l}a}
\affiliation{Institute of Physics, Polish Academy of Sciences,  PL-02 668 Warszawa, Poland}

\author{K. N. Trohidou}
\affiliation{Institute for Advanced Materials, Physicochemical Processes, Nanotechnology \& Microsystems, NCSR Demokritos, GR-15310 Athens, Greece}

\author{A. Bonanni}
\affiliation{Institut f\"ur Halbleiter- und Festk\"orperphysik, Johannes Kepler University,  A-4040 Linz, Austria}

\author{D. Hommel}
\affiliation{Institute of Solid State Physics, University of Bremen, D-28359 Bremen, Germany}

\author{M. Sawicki}
\affiliation{Institute of Physics, Polish Academy of Sciences, PL-02 668 Warszawa, Poland}

\author{T. Dietl}
\email{dietl@ifpan.edu.pl}
\affiliation{Institute of Physics, Polish Academy of Sciences, PL-02 668 Warszawa, Poland}
\affiliation{Institute of Theoretical Physics, Faculty of Physics, University of Warsaw, PL-00 681 Warszawa, Poland}
\affiliation{WPI-Advanced Institute for Materials Research, Tohoku University, Sendai 980-8577, Japan}

\begin{abstract}
Molecular beam epitaxy has been employed to obtain Ga$_{1-x}$Mn$_x$N films with $x$ up to 10\% and Curie temperatures $T_{\text{C}}$ up to 13~K. The magnitudes of $T_{\text{C}}$ and their dependence on $x$, $T_{\text{C}}(x) \propto x^m$, where $m = 2.2 \pm 0.2$ are quantitatively described by a tight binding model of superexchange interactions and Monte Carlo simulations of $T_{\text{C}}$. The critical behavior of this dilute magnetic insulator shows strong deviations from the magnetically clean case ($x = 1$), in particular, (i) an apparent breakdown of the Harris criterion; (ii) a non-monotonic crossover in the values of the susceptibility critical exponent $\gamma_{\text{eff}}$ between the high temperature and critical regimes, and (iii) a smearing of the critical region, which can be explained either by the Griffiths effects or by macroscopic inhomogeneities in the spin distribution with a variance $\Delta x = (0.2 \pm 0.1)$\%.
\end{abstract}

\date{\today}

\maketitle


Over the last 15 years, Ga$_{1-x}$Mn$_x$As and related dilute magnetic semiconductors (DMSs), such as In$_{1-x}$Mn$_x$As,\cite{Ohno:1998_S} have reached the status of model hole-mediated ferromagnetic systems\cite{Jungwirth:2006_RMP,MacDonald:2005_NM} in which a range of novel phenomena and functionalities have been demonstrated,\cite{Dietl:2010_NM} and transferred to ferromagnetic metals.\cite{Ohno:2010_NM} Surprisingly, however, the presence of ferromagnetic interactions was also detected in Ga$_{1-x}$Mn$_x$N,\cite{Kondo:2002_JCG,Sarigiannidou:2006_PRB,Freeman:2007_PRB,Bonanni:2011_PRB,Sawicki:2012_PRB} despite that in this compound the Fermi level is pinned in the mid-gap region, precluding the existence of carrier-mediated spin-spin coupling.  It was suggested\cite{Bonanni:2011_PRB,Sawicki:2012_PRB} that this puzzle can be resolved by noting that for Mn$^{3+}$ ions, the short-range superexchange acquires a ferromagnetic character, as found theoretically for tetrahedrally coordinated magnetic cations with partly filled $t_2$ orbitals, such as Cr$^{2+}$ in II-VI compounds.\cite{Blinowski:1996_PRB} Ferromagnetic coupling in such systems was also implied by {\em ab initio} studies, whose results were interpreted in terms of double exchange.\cite{Sato:2010_RMP}  The family of dilute ferromagnetic insulators is, actually, much wider, and contains also ferromagnetic topological insulators, including  Cr$_{1-x}$(Bi$_y$Sb$_{1-y}$)$_{2-x}$Te$_3$,\cite{Chang:2013_AM,Chang:2013_S} whose ferromagnetism was assigned to interband spin polarization.\cite{Yu:2010_S}

As shown recently, molecular beam epitaxy (MBE) allows one to obtain Ga$_{1-x}$Mn$_x$N films with Mn content  $x$ reaching 10\%, in which, due to the high cation density and the absence of competing antiferromagnetic interactions, the magnitude of the saturation magnetization at $\sim$70~kOe exceeds those reported to date for any other DMSs.\cite{Kunert:2012_APL} This progress indicates that Ga$_{1-x}$Mn$_x$N emerges as a model system making it possible to explore properties and functionalities specific to dilute ferromagnetic insulators.

Here we present detailed magnetization studies for MBE-grown films thoroughly characterized by a number of structure-sensitive and element-specific methods.
We show that the dependence of the Curie temperature $T_{\text{C}}$ on $x$ is, for such samples, in a quantitative agreement with theoretical results obtained by us combining a tight-binding evaluation of the exchange integrals for short-range ferromagnetic superexchange with Monte Carlo simulations of $T_{\text{C}}$. Having in hand the system with short-range ferromagnetic interactions between randomly distributed localized spins, we address experimentally the fundamental and long standing question on how disorder influences the critical behavior of  continuous phase transitions. Our results confirm experimentally that in contrast to the magnetically clean case ($x =1$), for which there is a monotonic crossover between the mean-field and the renormalization-group (RG) value of the effective critical exponent $\gamma_{\text{eff}}$ on approaching $T_{\text{C}}^+$, in the alloys studied here $\gamma_{\text{eff}}$ goes through a maximum, a behavior anticipated by a RG theory\cite{Sobotta:1982_JMMM,Dudka:2003_JMMM} and confirmed recently by massive Monte Carlo simulations\cite{Priour:2010_PRB} as well as showing up in experiments on metallic alloys.\cite{Fahnle:1983_JMMM,Babu:1997_JPCM,Perumal:2001_JMMM} Surprisingly, however, our results close to $T_{\text{C}}$ point to a certain smearing of the transition, which  may result from macroscopic inhomogeneities in the Mn content but which is also anticipated within the Griffiths scenario.\cite{Griffiths:1969_PRL,Galitski:2004_PRL}


Single crystalline layers of Ga$_{1-x}$Mn$_x$N, for which magnetic data are reported here, have been deposited by MBE at the substrate temperature $T_{\text{s}} = 730$ or $760^{\text{o}}$C under nominally nitrogen-rich growth-conditions on templates consisting of 2~$\mu$m GaN(0001) grown by metal-organic vapor-phase epitaxy on $c$-plane sapphire.\cite{Kunert:2012_APL}  Two samples ($x = 6.5$\% and 9.5\%) have been grown in the presence of Si, whose concentration according to secondary ion mass spectroscopy (SIMS) is more than two orders of magnitude lower than the one of Mn. In all films we do not detect any traces of Mn aggregation by high-resolution transmission electron microscopy (HRTEM). The thickness of the layers lies between 150 and 250~nm, and has been cross-checked by reflectometry, scanning electron microscopy (SEM), and SIMS.
Furthermore, for these samples, the Mn concentration $x$ evaluated by Rutherford backscattering spectrometry (RBS) and SIMS (calibrated by electron microprobe using a 1~$\mu$ thick (Ga,Mn)N film) agree within $\Delta x =2$\% with the values $x_{\text{eff}}$ determined from the magnitude of the magnetization at 1.85~K and in 70~kOe.\cite{Kunert:2012_APL} We assume the Mn magnetic moment to be $3.72\mu_{\text{B}}$, as obtained for Ga-substitutional Mn$^{3+}$ ions in the high $S=2$ spin state in GaN.\cite{Gosk:2005_PRB,Stefanowicz:2010_PRBa}
This generally good agreement, $x_{\text{eff}} \simeq x$, points to a relatively small concentration of Mn$^{2+}$ ions for which antiferromagnetic interactions result in $x_{\text{eff}} < x$.\cite{Granville:2010_PRB}

\begin{figure}[t]
		\centering
        \includegraphics[width=8.4 cm]{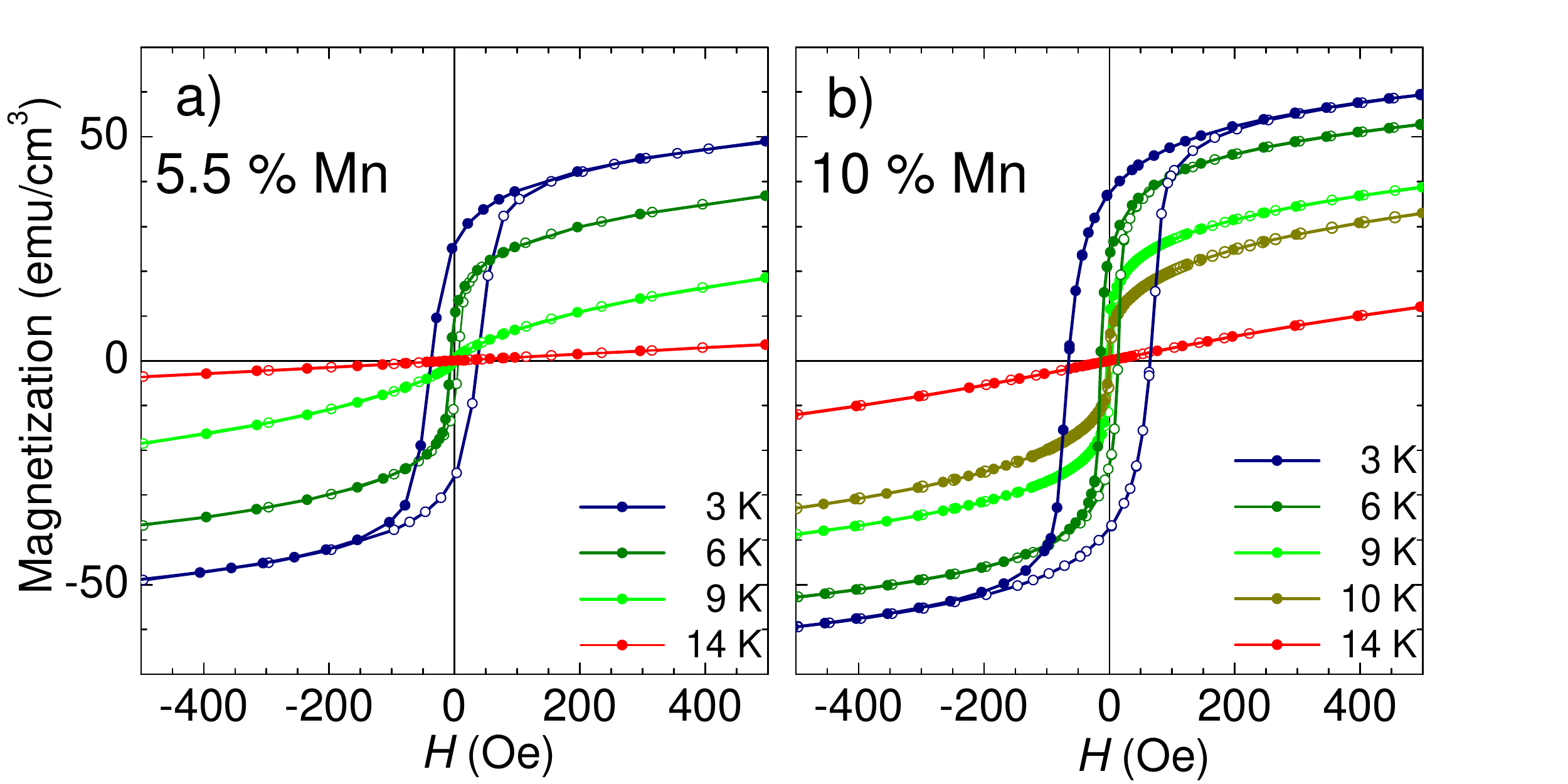}
      \caption{(Color online)  Magnetic hysteresis at selected temperatures for two Ga$_{1-x}$Mn$_x$N samples with Mn content $x$ = 5.5\% in panel (a) and 10\% in panel (b).}
     \label{fig:Hysteresis}
\end{figure}
For magnetic measurements, the samples are cut to approximately $5\times5$ mm$^2$ specimens and washed in concentrated HCl to remove possible traces of ferrous contaminants from surfaces and edges. The measurements are performed in a Quantum Design MPMS 7~T SQUID magnetometer following strictly  the guidelines of precise magnetometry of thin layers on a substrate, as outlined recently.\cite{Sawicki:2011_SST} In particular, the absolute values of the Ga$_{1-x}$Mn$_x$N layers' moments are obtained after the substraction of a reference signal measured for a GaN layer grown and processed in the same way as Ga$_{1-x}$Mn$_x$N samples, the reference signal being scaled according to the sample and the reference weights. The SQUID scaling factors dependent on the shape of the specimens are also incorporated into the procedure.\cite{Sawicki:2011_SST} The data sets are collected for both in-plane ($H \perp c$) and perpendicular ($H \parallel c$) alignments of the samples' face with respect to the external magnetic field, and thus to the axis of the SQUID detection coils. The low-field data indicate the existence of a sizable easy-plane magnetic anisotropy, so that we discuss results obtained for the in-plane magnetic field.

For studies in low magnetic fields, the samples are cooled down at $H  =1$~kOe.  Then the field is quenched, using the magnet reset option, down to $\sim 80$~mOe, as assessed by the magnetic moment of Dy$_2$O$_3$ paramagnetic salt. Under these conditions, the thermoremanent moment (TRM) is collected on increasing temperature until the TRM drops to zero. This is usually followed immediately by a magnetic moment measurement on \emph{decreasing} temperature at exactly the same zero-field conditions. The zero-field-cooled (0FC) values yield both direct information on the temperature at which the long-range-coupled spontaneous moment is created and an assessment of the magnitude of this moment.\cite{Sawicki:2010_NP}
The same low field conditions are set to study both the low field ($-20 < H < 100$~Oe) magnetic isotherms and the AC magnetic susceptibility.


\begin{figure}[t]
  \centering
  \includegraphics[width=8.5cm]{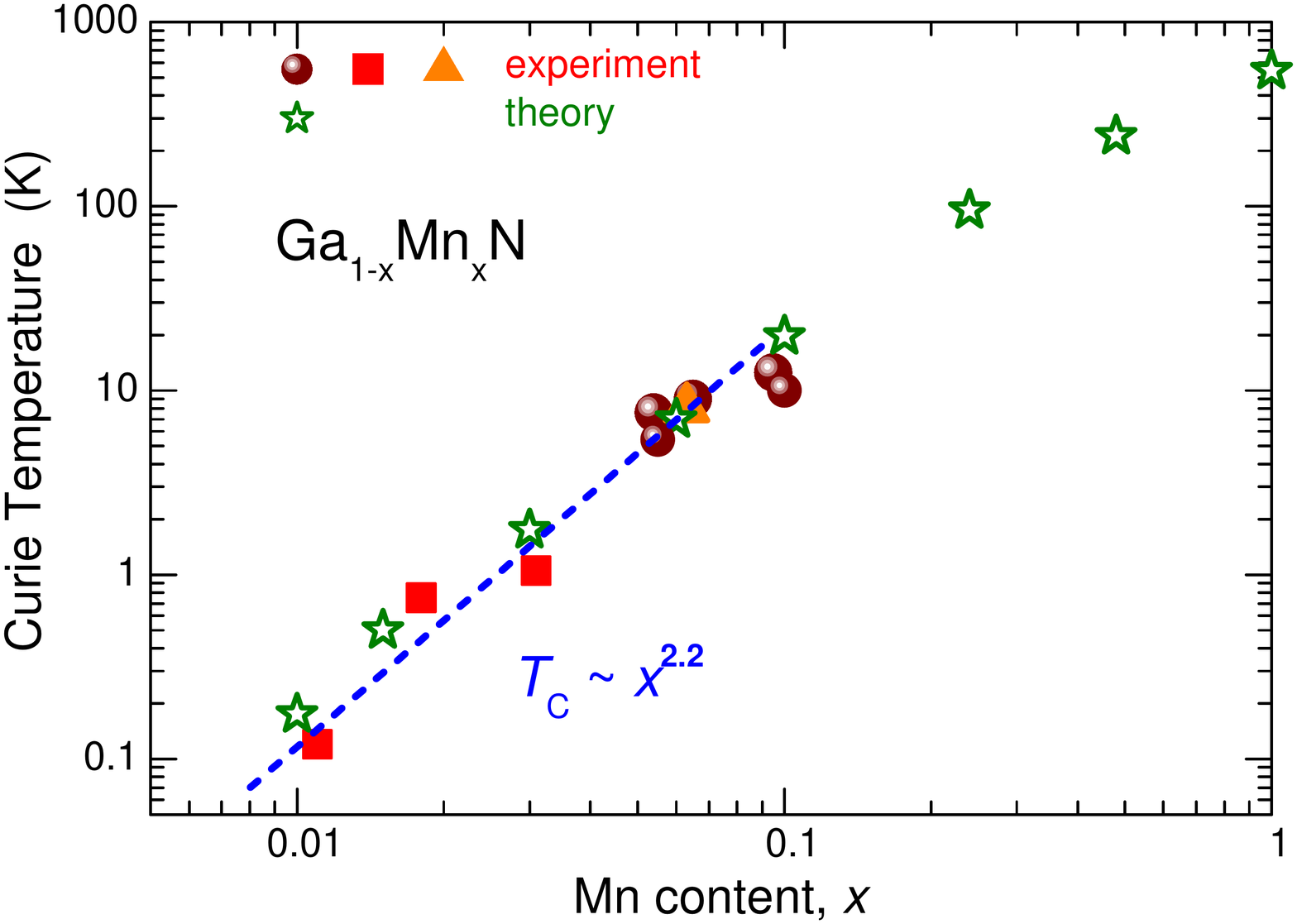}
  \caption{(Color online)  Experimental Curie temperatures as a function of Mn content $x$ (circles), together with the experimental result of Refs.~\onlinecite{Sawicki:2012_PRB} and \onlinecite{Sarigiannidou:2006_PRB} (squares and triangle, respectively). The dotted line indicates the scaling dependence  $T_{\text{C}} \propto x^{m}$ with $m=2.2$.  Results of  Monte Carlo simulations with exchange integrals from the tight binding model\cite{Sawicki:2012_PRB} (stars) are also shown.}
  \label{fig:phase_diagram}
\end{figure}
As shown in Fig.~\ref{fig:Hysteresis}, open magnetic hysteresis loops are observed at low temperatures, and their coercivity increases with $x$. As previously,\cite{Sawicki:2012_PRB} we identify the hysteresis onsets with $T_{\text{C}}$. According to Fig.~\ref{fig:phase_diagram}, the new values of $T_{\text{C}}$ confirm the trend, $T_{\text{C}} \propto x^{m}$, where $m = 2.2 \pm 0.2$. This supports the superexchange scenario,\cite{Bonanni:2011_PRB,Sawicki:2012_PRB} as  the same value of $m$ describes the dependence of spin-glass freezing temperatures on $x$ in Mn- and Co-doped DMSs,\cite{Twardowski:1987_PRB,Swagten:1992_PRB,Sawicki:2012_arXiv} in which the antiferromagnetic superexchange is an established spin coupling mechanism.  Moreover, as seen in Fig.~\ref{fig:phase_diagram}, the experimental results are in a remarkable agreement with the $T_{\text{C}}$ values obtained from the tight-binding and Monte Carlo simulations of ferromagnetic superexchange between Mn$^{3+}$ ions in zinc-blende GaN.\cite{Blinowski:1996_PRB,Sawicki:2012_PRB} In comparison with to our previous theoretical model,\cite{Sawicki:2012_PRB} we take now into account Mn-Mn exchange energies $J_{ij}$ up to the 16$^{th}$ cation coordination sphere, which allows us computing $T_{\text{C}}$ down to $x = 1$\%. Furthermore, confirming the previous suggestion,\cite{Sawicki:2012_PRB} we find a better agreement between theoretical and experimental $T_{\text{C}}$ values after changing the magnitude of the charge transfer parameter $e_2$ from 4.8 to 4.4~eV, {\em i.e.}, within its expected experimental uncertainty.

\begin{figure}[t]
		\centering
        \includegraphics[width=8 cm] {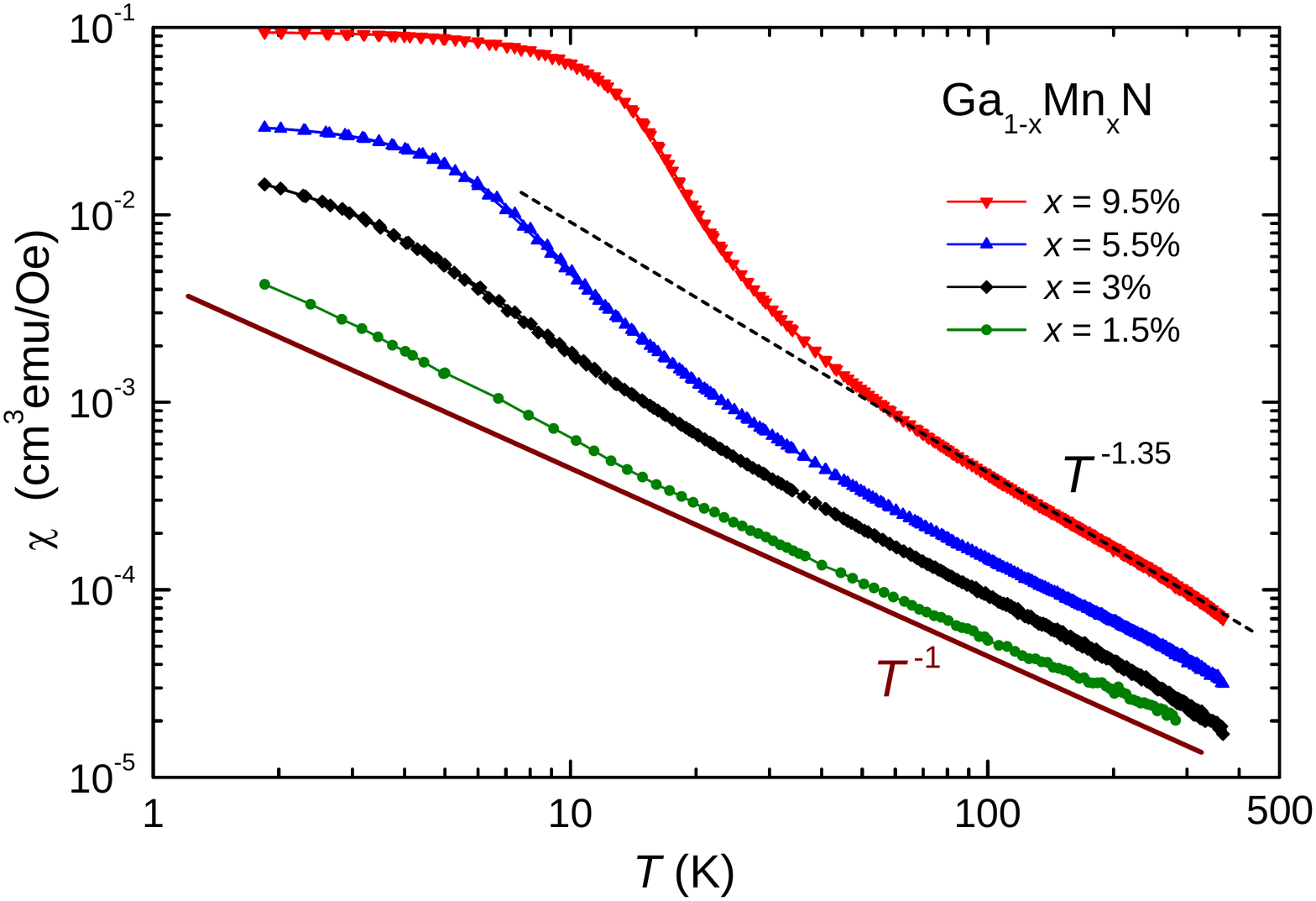}
      \caption{(Color online)   Temperature dependence of the magnetic susceptibility for selected Ga$_{1-x}$Mn$_x$N layers at 1~kOe.  The thick solid line traces the Curie law $\chi(T) \propto 1/T$. The dotted line indicates that at high temperatures $\chi(T) \propto T^{-\alpha}$, where $\alpha > 1$, a dependence specific for random ferromagnets with short range spin-spin interactions.}
     \label{fig:ChiLogLog}
\end{figure}

Having determined the origin and the range of spin-spin coupling we focus on the critical characteristics. They demonstrate striking differences compared to magnetically clean systems ($x = 1$) despite that the Harris criterion is fulfilled for the universality class in question (the three dimensional $XY$ or Heisenberg case), so that no effect of randomness on critical exponents is expected.\cite{Harris:1974_PRL} In particular, according to Fig.~\ref{fig:ChiLogLog}, the magnetic susceptibility $\chi(T)= M/H$ at $T \gg T_{\text{C}}$, instead of the Curie law, shows $\chi(T) \propto T^{-\alpha}$, where  $\alpha > 1$.  We assign this non-standard dependence to a gradual formation of coupled neighbor spin clusters (spin pairs, triads,...) on lowering temperature. 
This reasoning implies $\alpha < 1$ when the coupling is antiferromagnetic,\cite{Bhatt:1982_PRL} as indeed observed for Mn and Co-based II-VI DMSs.\cite{Dietl:1987_JJAP,Sawicki:2012_arXiv}

\begin{figure}[tb]
        \includegraphics[width=8.5 cm]{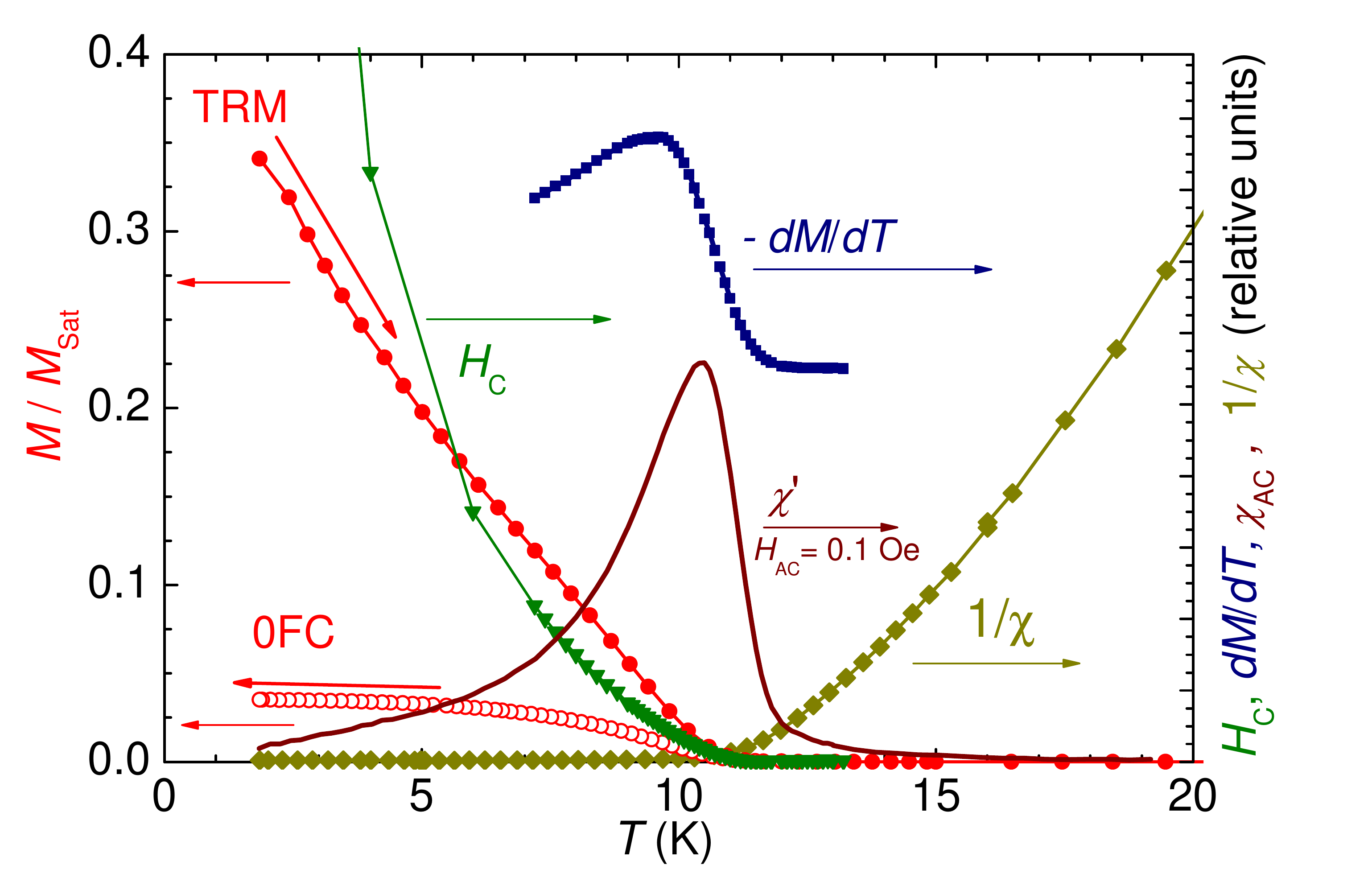}
      \caption{(Color online) Critical behavior of $x=10$\% Ga$_{1-x}$Mn$_x$N  and Curie point determination from the inverse ($1/\chi$) and AC ($\chi_{\text{AC}}$) magnetic susceptibility, coercive field ($H_{\text{c}}$), thermoremanent (TRM) and zero-field-cooled (0FC) magnetization $M$, as well as from the maximum of -d$M$/d$T$ at 0.3~Oe, the latter pointing to a lower value of $T_{\text{C}}$.}
\label{fig:Tc}
\end{figure}

According to the Griffiths suggestion,\cite{Griffiths:1969_PRL} the presence of preformed ferromagnetic clusters may smear the phase transition and shift up the apparent  value of $T_{\text{C}}$. A smearing may also result from macroscopic inhomogeneities of the Mn content $x$, which is an effect that is
always present in real alloys. According to the results summarized in Fig.~\ref{fig:Tc}, there is an excellent agreement between the values of $T_{\text{C}}$ determined from a maximum of the AC magnetic susceptibility $\chi_{\text{ac}}$ as well as from an extrapolation of the coercive field, $1/\chi$,  TRM, and 0FC magnitudes towards zero.
However, the position of the inflection point on $M(T, H=0.3$~Oe) points to a lower value of $T_{\text{C}}$ by about 1~K in all two samples for which such an analysis has been performed. This finding indicates a smearing of the transition as the former methods provide an upper bound of the $T_{\text{C}}$ distribution, in contrast to the inflection one that favors a statistically more representative lower bound.

\begin{figure}[tb]
        \includegraphics[width=8cm]{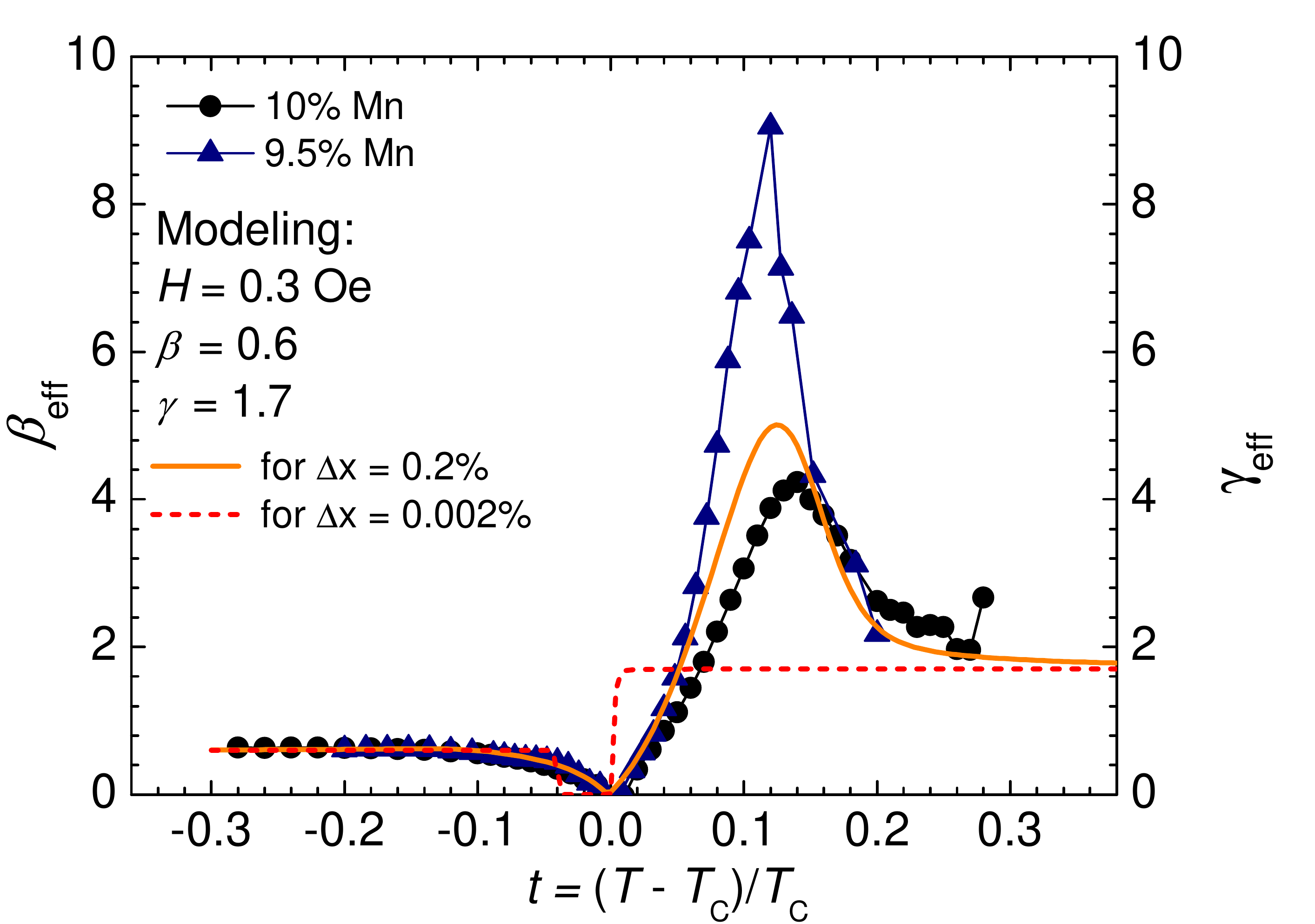}
      \caption{(Color online) Experimental (points) and theoretical (lines) temperature dependencies of the effective critical exponents for two Ga$_{1-x}$Mn$_x$N samples from different wafers. The values of $\beta$ and $\gamma$ as well two values of the variance $\Delta x$ in the Mn distribution employed in the calculations are displayed.}
\label{fig:exponents}
\end{figure}

Particularly informative in this context are the temperature dependencies of the effective critical exponents, $\beta_{\text{eff}} =  \mbox{d}\ln M/\mbox{d}\ln (-t)$ for $t < 0$  and $\gamma_{\text{eff}}  =  -\mbox{d}\ln M/\mbox{d}\ln t$ for $t>0$, where $M$ is established by polynomial interpolation of the experimental isotherms at 0.3 Oe and $t = (T-T_{\text{C}})/T_{\text{C}}$.   As seen in Fig.~\ref{fig:exponents}, $\gamma_{\text{eff}}(t)$ established for $H=0.3$~Oe  goes through a maximum, as expected theoretically.\cite{Dudka:2003_JMMM,Priour:2010_PRB} Surprisingly, however, instead of saturating on the RG values at $|t| \rightarrow 0$, $\beta_{\text{eff}}(t)$ and $\gamma_{\text{eff}}(t)$ vanish in this limit. We confirm that this behavior is not altered by the magnetic field up to 10~Oe and by the choice of $T_{\text{C}}$, used to evaluate $\beta_{\text{eff}}(t)$ and $\gamma_{\text{eff}}(t)$, within its lower and upper bounds.


We interpret these findings assuming a gaussian distribution ${\cal{P}}(x)$ of the Mn concentrations $x$ around the mean value $x_{\text{av}}$, so that $M_{\text{av}}(T)$ that serves to evaluate theoretically the magnitudes of $\beta_{\text{eff}}(t)$ and $\gamma_{\text{eff}}(t)$ is given by,
$M_{\text{av}} = \int \mbox{d}x {\cal{P}}(x)M(x)$,
where $M(x)=A_1(x)[T_{\text{C}}(x) -T]^{\beta}$ or $M(x)=A_2(x)[T-T_{\text{C}}(x)]^{-\gamma}$ with the constrain that $M(x)$ cannot be smaller or greater than $A_3(x)H^{1/\delta}$ for $t\leq 0$ and $t> 0$, respectively, and $T_{\text{C}}(x) =A_4x^m$. We treat the variance $\Delta x$ of ${\cal{P}}(x)$ as well as $\beta$ and $\gamma$ as fitting parameters. At the same time, we assume $m =2.2$ and $\delta = 4.8$, and we adjust the temperature independent constants $A_i$ to insure that the values of $T_{\text{C}}(x)$ and of $M(x)$ at $|t| \gg 0$ have the correct magnitudes. As seen in Fig.~\ref{fig:exponents}, the comparison of experimental and theoretical results then leads to $\Delta x =(0.2\pm0.1)$\%, $\beta = 0.6\pm0.1$, and $\gamma =1.7\pm0.1$. The magnitudes of $\beta$ and $\gamma$ are consistent with the theoretical expectations.\cite{Dudka:2003_JMMM,Priour:2010_PRB} It is to be seen whether  macroscopic inhomogeneities in the Mn distribution or rather the Griffiths effects account for the non-zero value of $\Delta x$ for the universality class and for interaction range in question.

In summary, our results, by extending significantly the concentrations range of Mn in GaN studied so far, support the view that ferromagnetic superexchange is the dominant coupling mechanism between Ga-substitutional Mn$^{3+}$ ions in Ga$_{1-x}$Mn$_x$N, leading to $T_{\text{C}} \simeq 12.5$~K at $x = 9.5$\%. According to our theoretical model room temperature ferromagnetism will appear for $x \gtrsim 50$\%, provided that no insulator-to-metal transition would shift the high $T_{\text{C}}$ regime to lower Mn contents.\cite{Dietl:2008_PRB} Detailed magnetization studies, particularly near $T_{\text{C}}$,  reemphasize the outstanding character of the critical behavior in random ferromagnets, in particular, an apparent breakdown of the Harris criterion, a non-monotonic crossover in the values of $\gamma_{\text{eff}}$ between the high temperature and critical regimes, and a smearing of the critical region either by the Griffiths effects or by macroscopic inhomogeneities in the spin distribution inherent to virtually all real magnetic alloys.

This work was supported by the FunDMS Advanced Grant of the ERC (Grant No.~227690) within the Ideas 7th Framework Programme of European Community,  by (Polish) National Science Centre through project MAESTRO  (Decision 2011/02/A/ST3/00125) and project PRELUDIUM (Decision 2012/07/N/ST3/03146), by the InTechFun (Grant No. POIG.01.03.01-00-159/08), and by the Austrian FWF (P22471, P20065 and P22477). Computer time in Athens was partly provided by the National Grid Infrastructure HellasGrid.


\begin{thebibliography}{36}%
\makeatletter
\providecommand \@ifxundefined [1]{%
 \@ifx{#1\undefined}
}%
\providecommand \@ifnum [1]{%
 \ifnum #1\expandafter \@firstoftwo
 \else \expandafter \@secondoftwo
 \fi
}%
\providecommand \@ifx [1]{%
 \ifx #1\expandafter \@firstoftwo
 \else \expandafter \@secondoftwo
 \fi
}%
\providecommand \natexlab [1]{#1}%
\providecommand \enquote  [1]{``#1''}%
\providecommand \bibnamefont  [1]{#1}%
\providecommand \bibfnamefont [1]{#1}%
\providecommand \citenamefont [1]{#1}%
\providecommand \href@noop [0]{\@secondoftwo}%
\providecommand \href [0]{\begingroup \@sanitize@url \@href}%
\providecommand \@href[1]{\@@startlink{#1}\@@href}%
\providecommand \@@href[1]{\endgroup#1\@@endlink}%
\providecommand \@sanitize@url [0]{\catcode `\\12\catcode `\$12\catcode
  `\&12\catcode `\#12\catcode `\^12\catcode `\_12\catcode `\%12\relax}%
\providecommand \@@startlink[1]{}%
\providecommand \@@endlink[0]{}%
\providecommand \url  [0]{\begingroup\@sanitize@url \@url }%
\providecommand \@url [1]{\endgroup\@href {#1}{\urlprefix }}%
\providecommand \urlprefix  [0]{URL }%
\providecommand \Eprint [0]{\href }%
\providecommand \doibase [0]{http://dx.doi.org/}%
\providecommand \selectlanguage [0]{\@gobble}%
\providecommand \bibinfo  [0]{\@secondoftwo}%
\providecommand \bibfield  [0]{\@secondoftwo}%
\providecommand \translation [1]{[#1]}%
\providecommand \BibitemOpen [0]{}%
\providecommand \bibitemStop [0]{}%
\providecommand \bibitemNoStop [0]{.\EOS\space}%
\providecommand \EOS [0]{\spacefactor3000\relax}%
\providecommand \BibitemShut  [1]{\csname bibitem#1\endcsname}%
\let\auto@bib@innerbib\@empty
\bibitem [{\citenamefont {Ohno}(1998)}]{Ohno:1998_S}%
  \BibitemOpen
  \bibfield  {author} {\bibinfo {author} {\bibfnamefont {H.}~\bibnamefont
  {Ohno}},\ }\href@noop {} {\bibfield  {journal} {\bibinfo  {journal}
  {Science}\ }\textbf {\bibinfo {volume} {281}},\ \bibinfo {pages} {951}
  (\bibinfo {year} {1998})}\BibitemShut {NoStop}%
\bibitem [{\citenamefont {Jungwirth}\ \emph {et~al.}(2006)\citenamefont
  {Jungwirth}, \citenamefont {Sinova}, \citenamefont {{Ma\v{s}ek}},
  \citenamefont {{Ku\v{c}era}},\ and\ \citenamefont
  {MacDonald}}]{Jungwirth:2006_RMP}%
  \BibitemOpen
  \bibfield  {author} {\bibinfo {author} {\bibfnamefont {T.}~\bibnamefont
  {Jungwirth}}, \bibinfo {author} {\bibfnamefont {J.}~\bibnamefont {Sinova}},
  \bibinfo {author} {\bibfnamefont {J.}~\bibnamefont {{Ma\v{s}ek}}}, \bibinfo
  {author} {\bibfnamefont {J.}~\bibnamefont {{Ku\v{c}era}}}, \ and\ \bibinfo
  {author} {\bibfnamefont {A.~H.}\ \bibnamefont {MacDonald}},\ }\href@noop {}
  {\bibfield  {journal} {\bibinfo  {journal} {Rev. Mod. Phys.}\ }\textbf
  {\bibinfo {volume} {78}},\ \bibinfo {pages} {809} (\bibinfo {year}
  {2006})}\BibitemShut {NoStop}%
\bibitem [{\citenamefont {MacDonald}\ \emph {et~al.}(2005)\citenamefont
  {MacDonald}, \citenamefont {Schiffer},\ and\ \citenamefont
  {Samarth}}]{MacDonald:2005_NM}%
  \BibitemOpen
  \bibfield  {author} {\bibinfo {author} {\bibfnamefont {A.~H.}\ \bibnamefont
  {MacDonald}}, \bibinfo {author} {\bibfnamefont {P.}~\bibnamefont {Schiffer}},
  \ and\ \bibinfo {author} {\bibfnamefont {N.}~\bibnamefont {Samarth}},\
  }\href@noop {} {\bibfield  {journal} {\bibinfo  {journal} {Nature Mater.}\
  }\textbf {\bibinfo {volume} {4}},\ \bibinfo {pages} {195} (\bibinfo {year}
  {2005})}\BibitemShut {NoStop}%
\bibitem [{\citenamefont {Dietl}(2010)}]{Dietl:2010_NM}%
  \BibitemOpen
  \bibfield  {author} {\bibinfo {author} {\bibfnamefont {T.}~\bibnamefont
  {Dietl}},\ }\href@noop {} {\bibfield  {journal} {\bibinfo  {journal} {Nature
  Mater.}\ }\textbf {\bibinfo {volume} {9}},\ \bibinfo {pages} {965} (\bibinfo
  {year} {2010})}\BibitemShut {NoStop}%
\bibitem [{\citenamefont {Ohno}(2010)}]{Ohno:2010_NM}%
  \BibitemOpen
  \bibfield  {author} {\bibinfo {author} {\bibfnamefont {H.}~\bibnamefont
  {Ohno}},\ }\href@noop {} {\bibfield  {journal} {\bibinfo  {journal} {Mat.
  Mater.}\ }\textbf {\bibinfo {volume} {9}},\ \bibinfo {pages} {952} (\bibinfo
  {year} {2010})}\BibitemShut {NoStop}%
\bibitem [{\citenamefont {Kondo}\ \emph {et~al.}(2002)\citenamefont {Kondo},
  \citenamefont {Kuwabara}, \citenamefont {Owa},\ and\ \citenamefont
  {Munekata}}]{Kondo:2002_JCG}%
  \BibitemOpen
  \bibfield  {author} {\bibinfo {author} {\bibfnamefont {T.}~\bibnamefont
  {Kondo}}, \bibinfo {author} {\bibfnamefont {S.}~\bibnamefont {Kuwabara}},
  \bibinfo {author} {\bibfnamefont {H.}~\bibnamefont {Owa}}, \ and\ \bibinfo
  {author} {\bibfnamefont {H.}~\bibnamefont {Munekata}},\ }\href@noop {}
  {\bibfield  {journal} {\bibinfo  {journal} {J. Cryst. Growth}\ }\textbf
  {\bibinfo {volume} {237}},\ \bibinfo {pages} {1353} (\bibinfo {year}
  {2002})}\BibitemShut {NoStop}%
\bibitem [{\citenamefont {Sarigiannidou}\ \emph {et~al.}(2006)\citenamefont
  {Sarigiannidou}, \citenamefont {Wilhelm}, \citenamefont {Monroy},
  \citenamefont {Galera}, \citenamefont {Bellet-Amalric}, \citenamefont
  {Rogalev}, \citenamefont {Goulon}, \citenamefont {Cibert},\ and\
  \citenamefont {Mariette}}]{Sarigiannidou:2006_PRB}%
  \BibitemOpen
  \bibfield  {author} {\bibinfo {author} {\bibfnamefont {E.}~\bibnamefont
  {Sarigiannidou}}, \bibinfo {author} {\bibfnamefont {F.}~\bibnamefont
  {Wilhelm}}, \bibinfo {author} {\bibfnamefont {E.}~\bibnamefont {Monroy}},
  \bibinfo {author} {\bibfnamefont {R.~M.}\ \bibnamefont {Galera}}, \bibinfo
  {author} {\bibfnamefont {E.}~\bibnamefont {Bellet-Amalric}}, \bibinfo
  {author} {\bibfnamefont {A.}~\bibnamefont {Rogalev}}, \bibinfo {author}
  {\bibfnamefont {J.}~\bibnamefont {Goulon}}, \bibinfo {author} {\bibfnamefont
  {J.}~\bibnamefont {Cibert}}, \ and\ \bibinfo {author} {\bibfnamefont
  {H.}~\bibnamefont {Mariette}},\ }\href@noop {} {\bibfield  {journal}
  {\bibinfo  {journal} {Phys. Rev. B}\ }\textbf {\bibinfo {volume} {74}},\
  \bibinfo {pages} {041306(R)} (\bibinfo {year} {2006})}\BibitemShut {NoStop}%
\bibitem [{\citenamefont {Freeman}\ \emph {et~al.}(2007)\citenamefont
  {Freeman}, \citenamefont {Edmonds}, \citenamefont {Farley}, \citenamefont
  {Novikov}, \citenamefont {Campion}, \citenamefont {Foxon}, \citenamefont
  {Gallagher}, \citenamefont {Sarigiannidou},\ and\ \citenamefont {van~der
  Laan}}]{Freeman:2007_PRB}%
  \BibitemOpen
  \bibfield  {author} {\bibinfo {author} {\bibfnamefont {A.~A.}\ \bibnamefont
  {Freeman}}, \bibinfo {author} {\bibfnamefont {K.~W.}\ \bibnamefont
  {Edmonds}}, \bibinfo {author} {\bibfnamefont {N.~R.~S.}\ \bibnamefont
  {Farley}}, \bibinfo {author} {\bibfnamefont {S.~V.}\ \bibnamefont {Novikov}},
  \bibinfo {author} {\bibfnamefont {R.~P.}\ \bibnamefont {Campion}}, \bibinfo
  {author} {\bibfnamefont {C.~T.}\ \bibnamefont {Foxon}}, \bibinfo {author}
  {\bibfnamefont {B.~L.}\ \bibnamefont {Gallagher}}, \bibinfo {author}
  {\bibfnamefont {E.}~\bibnamefont {Sarigiannidou}}, \ and\ \bibinfo {author}
  {\bibfnamefont {G.}~\bibnamefont {van~der Laan}},\ }\href@noop {} {\bibfield
  {journal} {\bibinfo  {journal} {Phys. Rev. B}\ }\textbf {\bibinfo {volume}
  {76}},\ \bibinfo {pages} {081201} (\bibinfo {year} {2007})}\BibitemShut
  {NoStop}%
\bibitem [{\citenamefont {Bonanni}\ \emph {et~al.}(2011)\citenamefont
  {Bonanni}, \citenamefont {Sawicki}, \citenamefont {Devillers}, \citenamefont
  {Stefanowicz}, \citenamefont {Faina}, \citenamefont {Li}, \citenamefont
  {Winkler}, \citenamefont {Sztenkiel}, \citenamefont {Navarro-Quezada},
  \citenamefont {Rovezzi}, \citenamefont {Jakie{\l}a}, \citenamefont {Grois},
  \citenamefont {Wegscheider}, \citenamefont {Jantsch}, \citenamefont
  {Suffczy{\'n}ski}, \citenamefont {D'Acapito}, \citenamefont {Meingast},
  \citenamefont {Kothleitner},\ and\ \citenamefont {Dietl}}]{Bonanni:2011_PRB}%
  \BibitemOpen
  \bibfield  {author} {\bibinfo {author} {\bibfnamefont {A.}~\bibnamefont
  {Bonanni}}, \bibinfo {author} {\bibfnamefont {M.}~\bibnamefont {Sawicki}},
  \bibinfo {author} {\bibfnamefont {T.}~\bibnamefont {Devillers}}, \bibinfo
  {author} {\bibfnamefont {W.}~\bibnamefont {Stefanowicz}}, \bibinfo {author}
  {\bibfnamefont {B.}~\bibnamefont {Faina}}, \bibinfo {author} {\bibfnamefont
  {T.}~\bibnamefont {Li}}, \bibinfo {author} {\bibfnamefont {T.~E.}\
  \bibnamefont {Winkler}}, \bibinfo {author} {\bibfnamefont {D.}~\bibnamefont
  {Sztenkiel}}, \bibinfo {author} {\bibfnamefont {A.}~\bibnamefont
  {Navarro-Quezada}}, \bibinfo {author} {\bibfnamefont {M.}~\bibnamefont
  {Rovezzi}}, \bibinfo {author} {\bibfnamefont {R.}~\bibnamefont {Jakie{\l}a}},
  \bibinfo {author} {\bibfnamefont {A.}~\bibnamefont {Grois}}, \bibinfo
  {author} {\bibfnamefont {M.}~\bibnamefont {Wegscheider}}, \bibinfo {author}
  {\bibfnamefont {W.}~\bibnamefont {Jantsch}}, \bibinfo {author} {\bibfnamefont
  {J.}~\bibnamefont {Suffczy{\'n}ski}}, \bibinfo {author} {\bibfnamefont
  {F.}~\bibnamefont {D'Acapito}}, \bibinfo {author} {\bibfnamefont
  {A.}~\bibnamefont {Meingast}}, \bibinfo {author} {\bibfnamefont
  {G.}~\bibnamefont {Kothleitner}}, \ and\ \bibinfo {author} {\bibfnamefont
  {T.}~\bibnamefont {Dietl}},\ }\href@noop {} {\bibfield  {journal} {\bibinfo
  {journal} {Phys. Rev. B}\ }\textbf {\bibinfo {volume} {84}},\ \bibinfo
  {pages} {035206} (\bibinfo {year} {2011})}\BibitemShut {NoStop}%
\bibitem [{\citenamefont {Sawicki}\ \emph
  {et~al.}(2012{\natexlab{a}})\citenamefont {Sawicki}, \citenamefont
  {Devillers}, \citenamefont {Ga{\l}\c{e}ski}, \citenamefont {Simserides},
  \citenamefont {Dobkowska}, \citenamefont {Faina}, \citenamefont {Grois},
  \citenamefont {Navarro-Quezada}, \citenamefont {Trohidou}, \citenamefont
  {Majewski}, \citenamefont {Dietl},\ and\ \citenamefont
  {Bonanni}}]{Sawicki:2012_PRB}%
  \BibitemOpen
  \bibfield  {author} {\bibinfo {author} {\bibfnamefont {M.}~\bibnamefont
  {Sawicki}}, \bibinfo {author} {\bibfnamefont {T.}~\bibnamefont {Devillers}},
  \bibinfo {author} {\bibfnamefont {S.}~\bibnamefont {Ga{\l}\c{e}ski}},
  \bibinfo {author} {\bibfnamefont {C.}~\bibnamefont {Simserides}}, \bibinfo
  {author} {\bibfnamefont {S.}~\bibnamefont {Dobkowska}}, \bibinfo {author}
  {\bibfnamefont {B.}~\bibnamefont {Faina}}, \bibinfo {author} {\bibfnamefont
  {A.}~\bibnamefont {Grois}}, \bibinfo {author} {\bibfnamefont
  {A.}~\bibnamefont {Navarro-Quezada}}, \bibinfo {author} {\bibfnamefont
  {K.~N.}\ \bibnamefont {Trohidou}}, \bibinfo {author} {\bibfnamefont {J.~A.}\
  \bibnamefont {Majewski}}, \bibinfo {author} {\bibfnamefont {T.}~\bibnamefont
  {Dietl}}, \ and\ \bibinfo {author} {\bibfnamefont {A.}~\bibnamefont
  {Bonanni}},\ }\href@noop {} {\bibfield  {journal} {\bibinfo  {journal} {Phys.
  Rev. B}\ }\textbf {\bibinfo {volume} {85}},\ \bibinfo {pages} {205204}
  (\bibinfo {year} {2012}{\natexlab{a}})}\BibitemShut {NoStop}%
\bibitem [{\citenamefont {Blinowski}\ \emph {et~al.}(1996)\citenamefont
  {Blinowski}, \citenamefont {Kacman},\ and\ \citenamefont
  {Majewski}}]{Blinowski:1996_PRB}%
  \BibitemOpen
  \bibfield  {author} {\bibinfo {author} {\bibfnamefont {J.}~\bibnamefont
  {Blinowski}}, \bibinfo {author} {\bibfnamefont {P.}~\bibnamefont {Kacman}}, \
  and\ \bibinfo {author} {\bibfnamefont {J.~A.}\ \bibnamefont {Majewski}},\
  }\href@noop {} {\bibfield  {journal} {\bibinfo  {journal} {Phys. Rev. B}\
  }\textbf {\bibinfo {volume} {53}},\ \bibinfo {pages} {9524} (\bibinfo {year}
  {1996})}\BibitemShut {NoStop}%
\bibitem [{\citenamefont {Sato}\ \emph {et~al.}(2010)\citenamefont {Sato},
  \citenamefont {Bergqvist}, \citenamefont {Kudrnovsk\'y}, \citenamefont
  {Dederichs}, \citenamefont {Eriksson}, \citenamefont {Turek}, \citenamefont
  {Sanyal}, \citenamefont {Bouzerar}, \citenamefont {Katayama-Yoshida},
  \citenamefont {Dinh}, \citenamefont {Fukushima}, \citenamefont {Kizaki},\
  and\ \citenamefont {Zeller}}]{Sato:2010_RMP}%
  \BibitemOpen
  \bibfield  {author} {\bibinfo {author} {\bibfnamefont {K.}~\bibnamefont
  {Sato}}, \bibinfo {author} {\bibfnamefont {L.}~\bibnamefont {Bergqvist}},
  \bibinfo {author} {\bibfnamefont {J.}~\bibnamefont {Kudrnovsk\'y}}, \bibinfo
  {author} {\bibfnamefont {P.~H.}\ \bibnamefont {Dederichs}}, \bibinfo {author}
  {\bibfnamefont {O.}~\bibnamefont {Eriksson}}, \bibinfo {author}
  {\bibfnamefont {I.}~\bibnamefont {Turek}}, \bibinfo {author} {\bibfnamefont
  {B.}~\bibnamefont {Sanyal}}, \bibinfo {author} {\bibfnamefont
  {G.}~\bibnamefont {Bouzerar}}, \bibinfo {author} {\bibfnamefont
  {H.}~\bibnamefont {Katayama-Yoshida}}, \bibinfo {author} {\bibfnamefont
  {V.~A.}\ \bibnamefont {Dinh}}, \bibinfo {author} {\bibfnamefont
  {T.}~\bibnamefont {Fukushima}}, \bibinfo {author} {\bibfnamefont
  {H.}~\bibnamefont {Kizaki}}, \ and\ \bibinfo {author} {\bibfnamefont
  {R.}~\bibnamefont {Zeller}},\ }\href@noop {} {\bibfield  {journal} {\bibinfo
  {journal} {Rev. Mod. Phys.}\ }\textbf {\bibinfo {volume} {82}},\ \bibinfo
  {pages} {1633} (\bibinfo {year} {2010})}\BibitemShut {NoStop}%
\bibitem [{\citenamefont {Chang}\ \emph
  {et~al.}(2013{\natexlab{a}})\citenamefont {Chang}, \citenamefont {Zhang},
  \citenamefont {Liu}, \citenamefont {Zhang}, \citenamefont {Feng},
  \citenamefont {Li}, \citenamefont {Wang}, \citenamefont {Chen}, \citenamefont
  {Dai}, \citenamefont {Fang}, \citenamefont {Qi}, \citenamefont {Zhang},
  \citenamefont {Wang}, \citenamefont {He}, \citenamefont {Ma},\ and\
  \citenamefont {Xue}}]{Chang:2013_AM}%
  \BibitemOpen
  \bibfield  {author} {\bibinfo {author} {\bibfnamefont {C.}~\bibnamefont
  {Chang}}, \bibinfo {author} {\bibfnamefont {J.}~\bibnamefont {Zhang}},
  \bibinfo {author} {\bibfnamefont {M.}~\bibnamefont {Liu}}, \bibinfo {author}
  {\bibfnamefont {Z.}~\bibnamefont {Zhang}}, \bibinfo {author} {\bibfnamefont
  {X.}~\bibnamefont {Feng}}, \bibinfo {author} {\bibfnamefont {K.}~\bibnamefont
  {Li}}, \bibinfo {author} {\bibfnamefont {L.}~\bibnamefont {Wang}}, \bibinfo
  {author} {\bibfnamefont {X.}~\bibnamefont {Chen}}, \bibinfo {author}
  {\bibfnamefont {X.}~\bibnamefont {Dai}}, \bibinfo {author} {\bibfnamefont
  {Z.}~\bibnamefont {Fang}}, \bibinfo {author} {\bibfnamefont {X.}~\bibnamefont
  {Qi}}, \bibinfo {author} {\bibfnamefont {S.}~\bibnamefont {Zhang}}, \bibinfo
  {author} {\bibfnamefont {Y.}~\bibnamefont {Wang}}, \bibinfo {author}
  {\bibfnamefont {K.}~\bibnamefont {He}}, \bibinfo {author} {\bibfnamefont
  {X.}~\bibnamefont {Ma}}, \ and\ \bibinfo {author} {\bibfnamefont
  {Q.}~\bibnamefont {Xue}},\ }\href@noop {} {\bibfield  {journal} {\bibinfo
  {journal} {Adv. Mater.}\ }\textbf {\bibinfo {volume} {25}},\ \bibinfo {pages}
  {1065} (\bibinfo {year} {2013}{\natexlab{a}})}\BibitemShut {NoStop}%
\bibitem [{\citenamefont {Chang}\ \emph
  {et~al.}(2013{\natexlab{b}})\citenamefont {Chang}, \citenamefont {Zhang},
  \citenamefont {Feng}, \citenamefont {Shen}, \citenamefont {Zhang},
  \citenamefont {Guo}, \citenamefont {Li}, \citenamefont {Ou}, \citenamefont
  {Wei}, \citenamefont {Wang}, \citenamefont {Ji}, \citenamefont {Feng},
  \citenamefont {Ji}, \citenamefont {Chen}, \citenamefont {Jia}, \citenamefont
  {Dai}, \citenamefont {Fang}, \citenamefont {Zhang}, \citenamefont {He},
  \citenamefont {Wang}, \citenamefont {Lu}, \citenamefont {Ma},\ and\
  \citenamefont {Xue}}]{Chang:2013_S}%
  \BibitemOpen
  \bibfield  {author} {\bibinfo {author} {\bibfnamefont {C.}~\bibnamefont
  {Chang}}, \bibinfo {author} {\bibfnamefont {J.}~\bibnamefont {Zhang}},
  \bibinfo {author} {\bibfnamefont {X.}~\bibnamefont {Feng}}, \bibinfo {author}
  {\bibfnamefont {J.}~\bibnamefont {Shen}}, \bibinfo {author} {\bibfnamefont
  {Z.}~\bibnamefont {Zhang}}, \bibinfo {author} {\bibfnamefont
  {M.}~\bibnamefont {Guo}}, \bibinfo {author} {\bibfnamefont {K.}~\bibnamefont
  {Li}}, \bibinfo {author} {\bibfnamefont {Y.}~\bibnamefont {Ou}}, \bibinfo
  {author} {\bibfnamefont {P.}~\bibnamefont {Wei}}, \bibinfo {author}
  {\bibfnamefont {L.}~\bibnamefont {Wang}}, \bibinfo {author} {\bibfnamefont
  {Z.}~\bibnamefont {Ji}}, \bibinfo {author} {\bibfnamefont {Y.}~\bibnamefont
  {Feng}}, \bibinfo {author} {\bibfnamefont {S.}~\bibnamefont {Ji}}, \bibinfo
  {author} {\bibfnamefont {X.}~\bibnamefont {Chen}}, \bibinfo {author}
  {\bibfnamefont {J.}~\bibnamefont {Jia}}, \bibinfo {author} {\bibfnamefont
  {X.}~\bibnamefont {Dai}}, \bibinfo {author} {\bibfnamefont {Z.}~\bibnamefont
  {Fang}}, \bibinfo {author} {\bibfnamefont {S.}~\bibnamefont {Zhang}},
  \bibinfo {author} {\bibfnamefont {K.}~\bibnamefont {He}}, \bibinfo {author}
  {\bibfnamefont {Y.}~\bibnamefont {Wang}}, \bibinfo {author} {\bibfnamefont
  {L.}~\bibnamefont {Lu}}, \bibinfo {author} {\bibfnamefont {X.}~\bibnamefont
  {Ma}}, \ and\ \bibinfo {author} {\bibfnamefont {Q.}~\bibnamefont {Xue}},\
  }\href@noop {} {\bibfield  {journal} {\bibinfo  {journal} {Science}\ }\textbf
  {\bibinfo {volume} {340}},\ \bibinfo {pages} {167} (\bibinfo {year}
  {2013}{\natexlab{b}})}\BibitemShut {NoStop}%
\bibitem [{\citenamefont {Yu}\ \emph {et~al.}(2010)\citenamefont {Yu},
  \citenamefont {Zhang}, \citenamefont {Zhang}, \citenamefont {Zhang},
  \citenamefont {Dai},\ and\ \citenamefont {Fang}}]{Yu:2010_S}%
  \BibitemOpen
  \bibfield  {author} {\bibinfo {author} {\bibfnamefont {R.}~\bibnamefont
  {Yu}}, \bibinfo {author} {\bibfnamefont {W.}~\bibnamefont {Zhang}}, \bibinfo
  {author} {\bibfnamefont {H.}~\bibnamefont {Zhang}}, \bibinfo {author}
  {\bibfnamefont {S.}~\bibnamefont {Zhang}}, \bibinfo {author} {\bibfnamefont
  {X.}~\bibnamefont {Dai}}, \ and\ \bibinfo {author} {\bibfnamefont
  {Z.}~\bibnamefont {Fang}},\ }\href@noop {} {\bibfield  {journal} {\bibinfo
  {journal} {Science}\ }\textbf {\bibinfo {volume} {329}},\ \bibinfo {pages}
  {61} (\bibinfo {year} {2010})}\BibitemShut {NoStop}%
\bibitem [{\citenamefont {Kunert}\ \emph {et~al.}(2012)\citenamefont {Kunert},
  \citenamefont {Dobkowska}, \citenamefont {Li}, \citenamefont {Reuther},
  \citenamefont {Kruse}, \citenamefont {Figge}, \citenamefont {Jakie{\l}a},
  \citenamefont {Bonanni}, \citenamefont {Grenzer}, \citenamefont
  {Stefanowicz}, \citenamefont {von Borany}, \citenamefont {Sawicki},
  \citenamefont {Dietl},\ and\ \citenamefont {Hommel}}]{Kunert:2012_APL}%
  \BibitemOpen
  \bibfield  {author} {\bibinfo {author} {\bibfnamefont {G.}~\bibnamefont
  {Kunert}}, \bibinfo {author} {\bibfnamefont {S.}~\bibnamefont {Dobkowska}},
  \bibinfo {author} {\bibfnamefont {T.}~\bibnamefont {Li}}, \bibinfo {author}
  {\bibfnamefont {H.}~\bibnamefont {Reuther}}, \bibinfo {author} {\bibfnamefont
  {C.}~\bibnamefont {Kruse}}, \bibinfo {author} {\bibfnamefont
  {S.}~\bibnamefont {Figge}}, \bibinfo {author} {\bibfnamefont
  {R.}~\bibnamefont {Jakie{\l}a}}, \bibinfo {author} {\bibfnamefont
  {A.}~\bibnamefont {Bonanni}}, \bibinfo {author} {\bibfnamefont
  {J.}~\bibnamefont {Grenzer}}, \bibinfo {author} {\bibfnamefont
  {W.}~\bibnamefont {Stefanowicz}}, \bibinfo {author} {\bibfnamefont
  {J.}~\bibnamefont {von Borany}}, \bibinfo {author} {\bibfnamefont
  {M.}~\bibnamefont {Sawicki}}, \bibinfo {author} {\bibfnamefont
  {T.}~\bibnamefont {Dietl}}, \ and\ \bibinfo {author} {\bibfnamefont
  {D.}~\bibnamefont {Hommel}},\ }\href@noop {} {\bibfield  {journal} {\bibinfo
  {journal} {Appl. Phys. Lett.}\ }\textbf {\bibinfo {volume} {101}},\ \bibinfo
  {pages} {022413} (\bibinfo {year} {2012})}\BibitemShut {NoStop}%
\bibitem [{\citenamefont {Sobotta}(1982)}]{Sobotta:1982_JMMM}%
  \BibitemOpen
  \bibfield  {author} {\bibinfo {author} {\bibfnamefont {G.}~\bibnamefont
  {Sobotta}},\ }\href@noop {} {\bibfield  {journal} {\bibinfo  {journal} {J.
  Magn. Magn. Mater.}\ }\textbf {\bibinfo {volume} {28}},\ \bibinfo {pages} {1}
  (\bibinfo {year} {1982})}\BibitemShut {NoStop}%
\bibitem [{\citenamefont {Dudka}\ \emph {et~al.}(2003)\citenamefont {Dudka},
  \citenamefont {Folk}, \citenamefont {Holovatch},\ and\ \citenamefont
  {Ivaneiko}}]{Dudka:2003_JMMM}%
  \BibitemOpen
  \bibfield  {author} {\bibinfo {author} {\bibfnamefont {M.}~\bibnamefont
  {Dudka}}, \bibinfo {author} {\bibfnamefont {R.}~\bibnamefont {Folk}},
  \bibinfo {author} {\bibfnamefont {Y.}~\bibnamefont {Holovatch}}, \ and\
  \bibinfo {author} {\bibfnamefont {D.}~\bibnamefont {Ivaneiko}},\ }\href@noop
  {} {\bibfield  {journal} {\bibinfo  {journal} {J. Magn. Magn. Mater.}\
  }\textbf {\bibinfo {volume} {256}},\ \bibinfo {pages} {243} (\bibinfo {year}
  {2003})}\BibitemShut {NoStop}%
\bibitem [{\citenamefont {Priour}\ and\ \citenamefont
  {Das~Sarma}(2010)}]{Priour:2010_PRB}%
  \BibitemOpen
  \bibfield  {author} {\bibinfo {author} {\bibfnamefont {D.~J.}\ \bibnamefont
  {Priour}}\ and\ \bibinfo {author} {\bibfnamefont {S.}~\bibnamefont
  {Das~Sarma}},\ }\href@noop {} {\bibfield  {journal} {\bibinfo  {journal}
  {Phys. Rev. B}\ }\textbf {\bibinfo {volume} {81}},\ \bibinfo {pages} {224403}
  (\bibinfo {year} {2010})}\BibitemShut {NoStop}%
\bibitem [{\citenamefont {F{\"a}hnle}\ \emph {et~al.}(1983)\citenamefont
  {F{\"a}hnle}, \citenamefont {Herzer}, \citenamefont {Kronm{\"u}ller},
  \citenamefont {Meyer}, \citenamefont {Saile},\ and\ \citenamefont
  {Egami}}]{Fahnle:1983_JMMM}%
  \BibitemOpen
  \bibfield  {author} {\bibinfo {author} {\bibfnamefont {M.}~\bibnamefont
  {F{\"a}hnle}}, \bibinfo {author} {\bibfnamefont {G.}~\bibnamefont {Herzer}},
  \bibinfo {author} {\bibfnamefont {H.}~\bibnamefont {Kronm{\"u}ller}},
  \bibinfo {author} {\bibfnamefont {R.}~\bibnamefont {Meyer}}, \bibinfo
  {author} {\bibfnamefont {M.}~\bibnamefont {Saile}}, \ and\ \bibinfo {author}
  {\bibfnamefont {T.}~\bibnamefont {Egami}},\ }\href@noop {} {\bibfield
  {journal} {\bibinfo  {journal} {J. Magn. Magn. Mater.}\ }\textbf {\bibinfo
  {volume} {38}},\ \bibinfo {pages} {240} (\bibinfo {year} {1983})}\BibitemShut
  {NoStop}%
\bibitem [{\citenamefont {Babu}\ and\ \citenamefont
  {Kaul}(1997)}]{Babu:1997_JPCM}%
  \BibitemOpen
  \bibfield  {author} {\bibinfo {author} {\bibfnamefont {P.~D.}\ \bibnamefont
  {Babu}}\ and\ \bibinfo {author} {\bibfnamefont {S.~N.}\ \bibnamefont
  {Kaul}},\ }\href@noop {} {\bibfield  {journal} {\bibinfo  {journal} {J. Phys.
  Condens. Matter}\ }\textbf {\bibinfo {volume} {9}},\ \bibinfo {pages} {7189}
  (\bibinfo {year} {1997})}\BibitemShut {NoStop}%
\bibitem [{\citenamefont {Perumal}\ \emph {et~al.}(2001)\citenamefont
  {Perumal}, \citenamefont {Srinivas}, \citenamefont {Kim}, \citenamefont {Yu},
  \citenamefont {Rao},\ and\ \citenamefont {Dunlap}}]{Perumal:2001_JMMM}%
  \BibitemOpen
  \bibfield  {author} {\bibinfo {author} {\bibfnamefont {A.}~\bibnamefont
  {Perumal}}, \bibinfo {author} {\bibfnamefont {V.}~\bibnamefont {Srinivas}},
  \bibinfo {author} {\bibfnamefont {K.}~\bibnamefont {Kim}}, \bibinfo {author}
  {\bibfnamefont {S.}~\bibnamefont {Yu}}, \bibinfo {author} {\bibfnamefont
  {V.}~\bibnamefont {Rao}}, \ and\ \bibinfo {author} {\bibfnamefont
  {R.}~\bibnamefont {Dunlap}},\ }\href@noop {} {\bibfield  {journal} {\bibinfo
  {journal} {J. Magn. Magn. Mater.}\ }\textbf {\bibinfo {volume} {233}},\
  \bibinfo {pages} {280} (\bibinfo {year} {2001})}\BibitemShut {NoStop}%
\bibitem [{\citenamefont {Griffiths}(1969)}]{Griffiths:1969_PRL}%
  \BibitemOpen
  \bibfield  {author} {\bibinfo {author} {\bibfnamefont {R.~B.}\ \bibnamefont
  {Griffiths}},\ }\href@noop {} {\bibfield  {journal} {\bibinfo  {journal}
  {Phys. Rev. Lett.}\ }\textbf {\bibinfo {volume} {23}},\ \bibinfo {pages} {17}
  (\bibinfo {year} {1969})}\BibitemShut {NoStop}%
\bibitem [{\citenamefont {Galitski}\ \emph {et~al.}(2003)\citenamefont
  {Galitski}, \citenamefont {Kaminski},\ and\ \citenamefont
  {Sarma}}]{Galitski:2004_PRL}%
  \BibitemOpen
  \bibfield  {author} {\bibinfo {author} {\bibfnamefont {V.~M.}\ \bibnamefont
  {Galitski}}, \bibinfo {author} {\bibfnamefont {A.}~\bibnamefont {Kaminski}},
  \ and\ \bibinfo {author} {\bibfnamefont {S.~D.}\ \bibnamefont {Sarma}},\
  }\href@noop {} {\bibfield  {journal} {\bibinfo  {journal} {Phys. Rev. Lett.}\
  }\textbf {\bibinfo {volume} {90}},\ \bibinfo {pages} {107202} (\bibinfo
  {year} {2003})}\BibitemShut {NoStop}%
\bibitem [{\citenamefont {Gosk}\ \emph {et~al.}(2005)\citenamefont {Gosk},
  \citenamefont {Zaj{\c{a}}c}, \citenamefont {Wo{\l}o{\'s}}, \citenamefont
  {Kami{\'n}ska}, \citenamefont {Twardowski}, \citenamefont {Grzegory},
  \citenamefont {Bockowski},\ and\ \citenamefont {Porowski}}]{Gosk:2005_PRB}%
  \BibitemOpen
  \bibfield  {author} {\bibinfo {author} {\bibfnamefont {J.}~\bibnamefont
  {Gosk}}, \bibinfo {author} {\bibfnamefont {M.}~\bibnamefont {Zaj{\c{a}}c}},
  \bibinfo {author} {\bibfnamefont {A.}~\bibnamefont {Wo{\l}o{\'s}}}, \bibinfo
  {author} {\bibfnamefont {M.}~\bibnamefont {Kami{\'n}ska}}, \bibinfo {author}
  {\bibfnamefont {A.}~\bibnamefont {Twardowski}}, \bibinfo {author}
  {\bibfnamefont {I.}~\bibnamefont {Grzegory}}, \bibinfo {author}
  {\bibfnamefont {M.}~\bibnamefont {Bockowski}}, \ and\ \bibinfo {author}
  {\bibfnamefont {S.}~\bibnamefont {Porowski}},\ }\href@noop {} {\bibfield
  {journal} {\bibinfo  {journal} {Phys. Rev. B}\ }\textbf {\bibinfo {volume}
  {71}},\ \bibinfo {pages} {094432} (\bibinfo {year} {2005})}\BibitemShut
  {NoStop}%
\bibitem [{\citenamefont {Stefanowicz}\ \emph {et~al.}(2010)\citenamefont
  {Stefanowicz}, \citenamefont {Sztenkiel}, \citenamefont {Faina},
  \citenamefont {Grois}, \citenamefont {Rovezzi}, \citenamefont {Devillers},
  \citenamefont {Navarro-Quezada}, \citenamefont {Li}, \citenamefont
  {Jakie{\l}a}, \citenamefont {Sawicki}, \citenamefont {Dietl},\ and\
  \citenamefont {Bonanni}}]{Stefanowicz:2010_PRBa}%
  \BibitemOpen
  \bibfield  {author} {\bibinfo {author} {\bibfnamefont {W.}~\bibnamefont
  {Stefanowicz}}, \bibinfo {author} {\bibfnamefont {D.}~\bibnamefont
  {Sztenkiel}}, \bibinfo {author} {\bibfnamefont {B.}~\bibnamefont {Faina}},
  \bibinfo {author} {\bibfnamefont {A.}~\bibnamefont {Grois}}, \bibinfo
  {author} {\bibfnamefont {M.}~\bibnamefont {Rovezzi}}, \bibinfo {author}
  {\bibfnamefont {T.}~\bibnamefont {Devillers}}, \bibinfo {author}
  {\bibfnamefont {A.}~\bibnamefont {Navarro-Quezada}}, \bibinfo {author}
  {\bibfnamefont {T.}~\bibnamefont {Li}}, \bibinfo {author} {\bibfnamefont
  {R.}~\bibnamefont {Jakie{\l}a}}, \bibinfo {author} {\bibfnamefont
  {M.}~\bibnamefont {Sawicki}}, \bibinfo {author} {\bibfnamefont
  {T.}~\bibnamefont {Dietl}}, \ and\ \bibinfo {author} {\bibfnamefont
  {A.}~\bibnamefont {Bonanni}},\ }\href@noop {} {\bibfield  {journal} {\bibinfo
   {journal} {Phys. Rev. B}\ }\textbf {\bibinfo {volume} {81}},\ \bibinfo
  {pages} {235210} (\bibinfo {year} {2010})}\BibitemShut {NoStop}%
\bibitem [{\citenamefont {Granville}\ \emph {et~al.}(2010)\citenamefont
  {Granville}, \citenamefont {Ruck}, \citenamefont {Budde}, \citenamefont
  {Trodahl},\ and\ \citenamefont {Williams}}]{Granville:2010_PRB}%
  \BibitemOpen
  \bibfield  {author} {\bibinfo {author} {\bibfnamefont {S.}~\bibnamefont
  {Granville}}, \bibinfo {author} {\bibfnamefont {B.~J.}\ \bibnamefont {Ruck}},
  \bibinfo {author} {\bibfnamefont {F.}~\bibnamefont {Budde}}, \bibinfo
  {author} {\bibfnamefont {H.~J.}\ \bibnamefont {Trodahl}}, \ and\ \bibinfo
  {author} {\bibfnamefont {G.~V.~M.}\ \bibnamefont {Williams}},\ }\href@noop {}
  {\bibfield  {journal} {\bibinfo  {journal} {Phys. Rev. B}\ }\textbf {\bibinfo
  {volume} {81}},\ \bibinfo {pages} {184425} (\bibinfo {year}
  {2010})}\BibitemShut {NoStop}%
\bibitem [{\citenamefont {Sawicki}\ \emph {et~al.}(2011)\citenamefont
  {Sawicki}, \citenamefont {Stefanowicz},\ and\ \citenamefont
  {Ney}}]{Sawicki:2011_SST}%
  \BibitemOpen
  \bibfield  {author} {\bibinfo {author} {\bibfnamefont {M.}~\bibnamefont
  {Sawicki}}, \bibinfo {author} {\bibfnamefont {W.}~\bibnamefont
  {Stefanowicz}}, \ and\ \bibinfo {author} {\bibfnamefont {A.}~\bibnamefont
  {Ney}},\ }\href@noop {} {\bibfield  {journal} {\bibinfo  {journal} {Semicon.
  Sci. Technol.}\ }\textbf {\bibinfo {volume} {26}},\ \bibinfo {pages} {064006}
  (\bibinfo {year} {2011})}\BibitemShut {NoStop}%
\bibitem [{\citenamefont {Sawicki}\ \emph {et~al.}(2010)\citenamefont
  {Sawicki}, \citenamefont {Chiba}, \citenamefont {Korbecka}, \citenamefont
  {Nishitani}, \citenamefont {Majewski}, \citenamefont {Matsukura},
  \citenamefont {Dietl},\ and\ \citenamefont {Ohno}}]{Sawicki:2010_NP}%
  \BibitemOpen
  \bibfield  {author} {\bibinfo {author} {\bibfnamefont {M.}~\bibnamefont
  {Sawicki}}, \bibinfo {author} {\bibfnamefont {D.}~\bibnamefont {Chiba}},
  \bibinfo {author} {\bibfnamefont {A.}~\bibnamefont {Korbecka}}, \bibinfo
  {author} {\bibfnamefont {Y.}~\bibnamefont {Nishitani}}, \bibinfo {author}
  {\bibfnamefont {J.~A.}\ \bibnamefont {Majewski}}, \bibinfo {author}
  {\bibfnamefont {F.}~\bibnamefont {Matsukura}}, \bibinfo {author}
  {\bibfnamefont {T.}~\bibnamefont {Dietl}}, \ and\ \bibinfo {author}
  {\bibfnamefont {H.}~\bibnamefont {Ohno}},\ }\href@noop {} {\bibfield
  {journal} {\bibinfo  {journal} {Nature Phys.}\ }\textbf {\bibinfo {volume}
  {6}},\ \bibinfo {pages} {22} (\bibinfo {year} {2010})}\BibitemShut {NoStop}%
\bibitem [{\citenamefont {Twardowski}\ \emph {et~al.}(1987)\citenamefont
  {Twardowski}, \citenamefont {Swagten}, \citenamefont {de~Jonge},\ and\
  \citenamefont {Demianiuk}}]{Twardowski:1987_PRB}%
  \BibitemOpen
  \bibfield  {author} {\bibinfo {author} {\bibfnamefont {A.}~\bibnamefont
  {Twardowski}}, \bibinfo {author} {\bibfnamefont {H.~J.~M.}\ \bibnamefont
  {Swagten}}, \bibinfo {author} {\bibfnamefont {W.~J.~M.}\ \bibnamefont
  {de~Jonge}}, \ and\ \bibinfo {author} {\bibfnamefont {M.}~\bibnamefont
  {Demianiuk}},\ }\href@noop {} {\bibfield  {journal} {\bibinfo  {journal}
  {Phys. Rev. B}\ }\textbf {\bibinfo {volume} {36}},\ \bibinfo {pages} {7013}
  (\bibinfo {year} {1987})}\BibitemShut {NoStop}%
\bibitem [{\citenamefont {Swagten}\ \emph {et~al.}(1992)\citenamefont
  {Swagten}, \citenamefont {Twardowski}, \citenamefont {Eggenkamp},\ and\
  \citenamefont {de~Jonge}}]{Swagten:1992_PRB}%
  \BibitemOpen
  \bibfield  {author} {\bibinfo {author} {\bibfnamefont {H.~J.~M.}\
  \bibnamefont {Swagten}}, \bibinfo {author} {\bibfnamefont {A.}~\bibnamefont
  {Twardowski}}, \bibinfo {author} {\bibfnamefont {P.~J.~T.}\ \bibnamefont
  {Eggenkamp}}, \ and\ \bibinfo {author} {\bibfnamefont {W.~J.~M.}\
  \bibnamefont {de~Jonge}},\ }\href@noop {} {\bibfield  {journal} {\bibinfo
  {journal} {Phys. Rev. B}\ }\textbf {\bibinfo {volume} {46}},\ \bibinfo
  {pages} {188} (\bibinfo {year} {1992})}\BibitemShut {NoStop}%
\bibitem [{\citenamefont {Sawicki}\ \emph
  {et~al.}(2012{\natexlab{b}})\citenamefont {Sawicki}, \citenamefont
  {Guziewicz}, \citenamefont {Lukasiewicz}, \citenamefont {Proselkov},
  \citenamefont {Kowalik}, \citenamefont {Lisowski}, \citenamefont {Dluzewski},
  \citenamefont {Wittlin}, \citenamefont {Jaworski}, \citenamefont {Wolska},
  \citenamefont {Paszkowicz}, \citenamefont {Jakiela}, \citenamefont
  {Witkowski}, \citenamefont {Wachnicki}, \citenamefont {Klepka}, \citenamefont
  {Luqu}, \citenamefont {Arvanitis}, \citenamefont {Sobczak}, \citenamefont
  {Krawczyk}, \citenamefont {Jablonski}, \citenamefont {Stefanowicz},
  \citenamefont {Sztenkiel}, \citenamefont {Godlewski},\ and\ \citenamefont
  {Dietl}}]{Sawicki:2012_arXiv}%
  \BibitemOpen
  \bibfield  {author} {\bibinfo {author} {\bibfnamefont {M.}~\bibnamefont
  {Sawicki}}, \bibinfo {author} {\bibfnamefont {E.}~\bibnamefont {Guziewicz}},
  \bibinfo {author} {\bibfnamefont {M.~I.}\ \bibnamefont {Lukasiewicz}},
  \bibinfo {author} {\bibfnamefont {O.}~\bibnamefont {Proselkov}}, \bibinfo
  {author} {\bibfnamefont {I.~A.}\ \bibnamefont {Kowalik}}, \bibinfo {author}
  {\bibfnamefont {W.}~\bibnamefont {Lisowski}}, \bibinfo {author}
  {\bibfnamefont {P.}~\bibnamefont {Dluzewski}}, \bibinfo {author}
  {\bibfnamefont {A.}~\bibnamefont {Wittlin}}, \bibinfo {author} {\bibfnamefont
  {M.}~\bibnamefont {Jaworski}}, \bibinfo {author} {\bibfnamefont
  {A.}~\bibnamefont {Wolska}}, \bibinfo {author} {\bibfnamefont
  {W.}~\bibnamefont {Paszkowicz}}, \bibinfo {author} {\bibfnamefont
  {R.}~\bibnamefont {Jakiela}}, \bibinfo {author} {\bibfnamefont {B.~S.}\
  \bibnamefont {Witkowski}}, \bibinfo {author} {\bibfnamefont {L.}~\bibnamefont
  {Wachnicki}}, \bibinfo {author} {\bibfnamefont {M.~T.}\ \bibnamefont
  {Klepka}}, \bibinfo {author} {\bibfnamefont {F.~J.}\ \bibnamefont {Luqu}},
  \bibinfo {author} {\bibfnamefont {D.}~\bibnamefont {Arvanitis}}, \bibinfo
  {author} {\bibfnamefont {J.~W.}\ \bibnamefont {Sobczak}}, \bibinfo {author}
  {\bibfnamefont {M.}~\bibnamefont {Krawczyk}}, \bibinfo {author}
  {\bibfnamefont {A.}~\bibnamefont {Jablonski}}, \bibinfo {author}
  {\bibfnamefont {W.}~\bibnamefont {Stefanowicz}}, \bibinfo {author}
  {\bibfnamefont {D.}~\bibnamefont {Sztenkiel}}, \bibinfo {author}
  {\bibfnamefont {M.}~\bibnamefont {Godlewski}}, \ and\ \bibinfo {author}
  {\bibfnamefont {T.}~\bibnamefont {Dietl}},\ }\href@noop {} {\bibfield
  {journal} {\bibinfo  {journal} {arXiv:1201.5268}\ } (\bibinfo {year}
  {2012}{\natexlab{b}})}\BibitemShut {NoStop}%
\bibitem [{\citenamefont {Harris}\ and\ \citenamefont
  {Lubensky}(1974)}]{Harris:1974_PRL}%
  \BibitemOpen
  \bibfield  {author} {\bibinfo {author} {\bibfnamefont {A.~B.}\ \bibnamefont
  {Harris}}\ and\ \bibinfo {author} {\bibfnamefont {T.~C.}\ \bibnamefont
  {Lubensky}},\ }\href@noop {} {\bibfield  {journal} {\bibinfo  {journal}
  {Phys. Rev. Lett.}\ }\textbf {\bibinfo {volume} {33}},\ \bibinfo {pages}
  {1540} (\bibinfo {year} {1974})}\BibitemShut {NoStop}%
\bibitem [{\citenamefont {Bhatt}\ and\ \citenamefont
  {Lee}(1982)}]{Bhatt:1982_PRL}%
  \BibitemOpen
  \bibfield  {author} {\bibinfo {author} {\bibfnamefont {R.~N.}\ \bibnamefont
  {Bhatt}}\ and\ \bibinfo {author} {\bibfnamefont {P.~A.}\ \bibnamefont
  {Lee}},\ }\href@noop {} {\bibfield  {journal} {\bibinfo  {journal} {Phys.
  Rev. Lett.}\ }\textbf {\bibinfo {volume} {48}},\ \bibinfo {pages} {344}
  (\bibinfo {year} {1982})}\BibitemShut {NoStop}%
\bibitem [{\citenamefont {Dietl}(1987)}]{Dietl:1987_JJAP}%
  \BibitemOpen
  \bibfield  {author} {\bibinfo {author} {\bibfnamefont {T.}~\bibnamefont
  {Dietl}},\ }\href@noop {} {\bibfield  {journal} {\bibinfo  {journal} {Jpn. J.
  Appl. Phys.}\ }\textbf {\bibinfo {volume} {26S3}},\ \bibinfo {pages} {1907}
  (\bibinfo {year} {1987})}\BibitemShut {NoStop}%
\bibitem [{\citenamefont {Dietl}(2008)}]{Dietl:2008_PRB}%
  \BibitemOpen
  \bibfield  {author} {\bibinfo {author} {\bibfnamefont {T.}~\bibnamefont
  {Dietl}},\ }\href@noop {} {\bibfield  {journal} {\bibinfo  {journal} {Phys.
  Rev. B}\ }\textbf {\bibinfo {volume} {77}},\ \bibinfo {pages} {085208}
  (\bibinfo {year} {2008})}\BibitemShut {NoStop}%
\end{thebibliography}
%

\end{document}